\documentclass[reprint,aip,superscriptaddress]{revtex4-1}
\usepackage{siunitx}
\usepackage[utf8]{inputenc}
\usepackage[T1]{fontenc}
\usepackage{xcolor}
\usepackage{soul}
\usepackage{braket}
\usepackage{lineno}
\usepackage[linktocpage=true]{hyperref}
\hypersetup{colorlinks=true,citecolor=red,urlcolor=blue}
\usepackage{amsmath}
\usepackage{amssymb}
\usepackage{braket}
\usepackage{graphicx}
\usepackage{dcolumn}
\usepackage{bm}
\usepackage{gensymb}
\usepackage{multirow}

\usepackage{changes}
\usepackage{xcolor,soul}
\sethlcolor{red}

\begin{document}
\preprint{}
\title[ ]{Neutrino mass and nature through its mediation in atomic clock interference}  
\author{Jos\'e Bernabeu}\email{jose.bernabeu@uv.es}
\affiliation{Department of Theoretical Physics,University of Valencia and IFIC, Univ. Valencia - CSIC, Burjassot, Valencia E-46100 Spain}
\author{Dylan O. Sabulsky}\email{dylan.sabulsky@obspm.fr}
\affiliation{LNE-SYRTE, Observatoire de Paris–Universit\'{e} PSL, CNRS, Sorbonne Universit\'{e}, 61 Avenue de l’Observatoire, Paris F-75014, France}
\author{Federico S\'anchez}\email{federico.sancheznieto@unige.ch}
\affiliation{Département de Physique Nucléaire et Corpusculaire (DPNC), Université de Genève - Faculté des Sciences, Genève 4 CH-1211, Switzerland}
\author{Alejandro Segarra}\affiliation{Department of Theoretical Physics,University of Valencia and IFIC, Univ. Valencia - CSIC, Burjassot, Valencia E-46100 Spain}

\begin{abstract}
The absolute mass of neutrinos and their nature are presently unknown. 
Aggregate matter has a coherent weak charge leading to a repulsive interaction mediated by a neutrino pair. 
The virtual neutrinos are non-relativistic at micron distances, giving a distinct behavior for Dirac versus Majorana mass terms. 
Through the magnitude and the distance dependence of this effective potential allow for the disentanglement the Dirac or Majorana nature of the neutrino.
We propose an experiment to search for this potential based on the concept that the density dependent interaction of an atomic probe with a material source in one arm of an atomic clock interferometer generates a differential phase. 
The appropriate geometry of the device is selected using the saturation of the weak potential as a guide. 
The proposed experiment has the added benefit of being sensitive to gravity at micron distances.
A strategy to suppress the competing Casimir-Polder interaction, depending on the electronic structure of the material source, as well as a way to compensate the gravitational interaction in the two arms of the interferometer is discussed. 
\end{abstract}
\date{\today} 
\maketitle
\section{Introduction}
\par Neutrino flavor oscillations \cite{PhysRevLett.81.1562} prove that neutrinos have mass and mixing among the three flavors.
The aim of future oscillation experiments is the search for CP Violation in the lepton sector as well as the mass ordering of neutrino states via propagation in matter\cite{protocollaboration2018hyperkamiokande,dunecollaboration2016longbaseline}.
These experiments are \textit{blind} to fundamental neutrino properties, like the absolute scale of neutrino mass and the nature of massive neutrinos, be it either Dirac or Majorana.
The aforementioned neutrino properties are being studied, respectively, in the KATRIN experiment \cite{PhysRevLett.123.221802} through tritium beta decay and in a plethora of neutrinoless double beta decay experiments \cite{PhysRevLett.120.132503,PhysRevLett.124.122501} with improving sensitivity to this $\Delta L = 2$ transition, where $L$ is the global lepton number.
As neutrino masses have a scale below 1 eV, the difficulty of these searches at nuclear release energies, or in other proposed X-ray transitions like neutrinoless double electron capture \cite{BERNABEU198315} where the stimulated enhancement has been considered\cite{Bern2018}, is almost insurmountable.
\begin{figure}
\centering
\includegraphics[width=\linewidth]{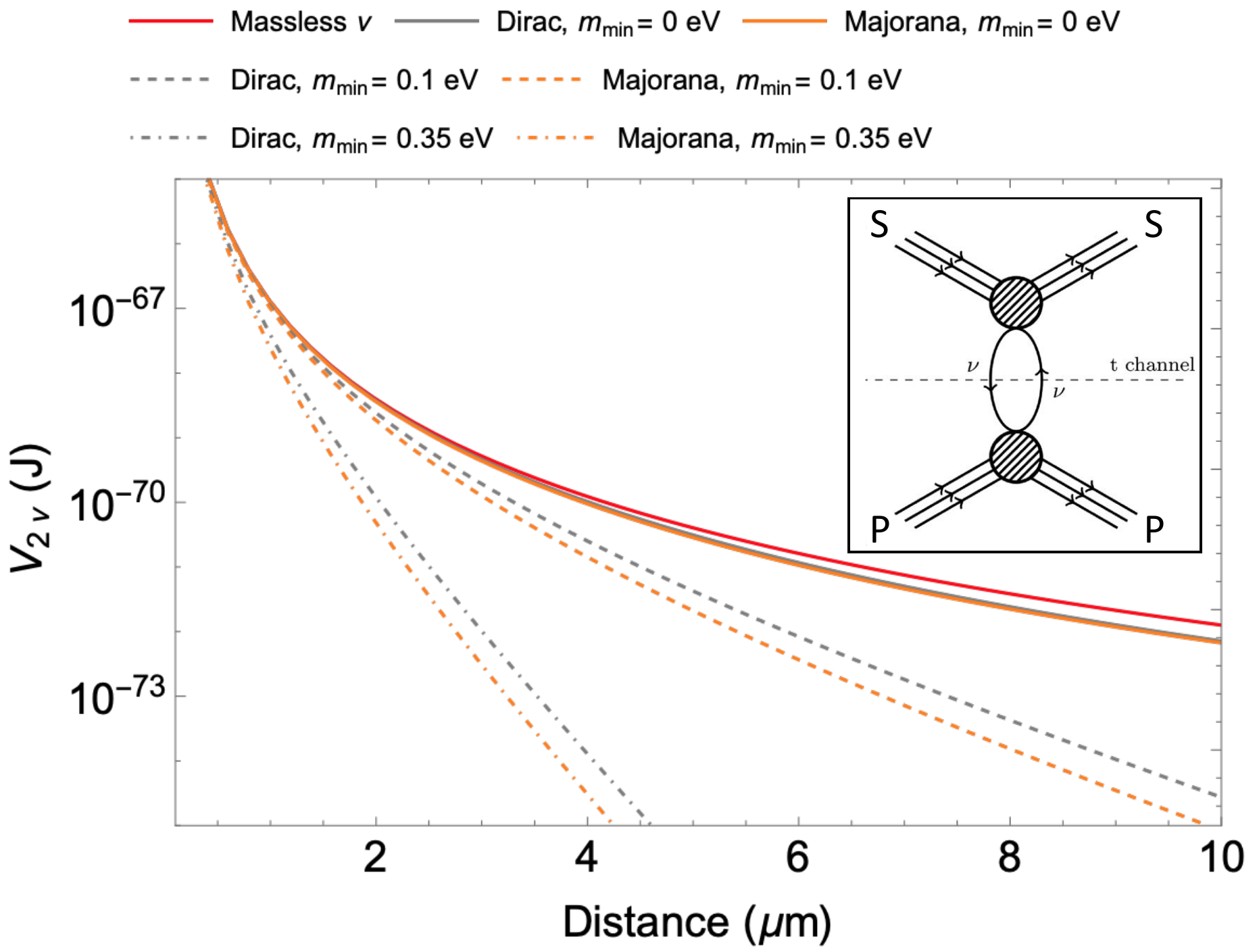}
\caption{Two neutrino exchange potential $V_{2\nu}(R)$\cite{PhysRevD.101.093004} between two atoms, $S \rightarrow {_{32}^{74}}\text{Ge}$ and $P \rightarrow {_{70}^{174}}\text{Yb}$.
We show several relevant cases. 
The massless neutrino limit where all neutrinos have null mass (red) and the scenarios for the minimum mass $m_{\text{min}}=(0,0.1,0.35)$ for both Dirac (gray; solid, dashed, and dot-dashed, respectively) and Majorana (orange; solid, dashed, and dot-dashed, respectively) neutrino models. 
The highest mass scenario we have selected (0.35~eV) is compatible with the final sensitivity of the direct mass neutrino measurement experiment KATRIN \cite{PhysRevLett.123.221802}. 
\label{fig:1}}
\end{figure}
\par These neutrino mass dependencies become apparent for non-relativistic neutrinos - this is to say the Dirac or Majorana nature becomes distinguishable through the mass terms. 
Is it possible to descend, experimentally, to energy scales where neutrinos are non-relativistic?
The only source existing in nature for such low energy neutrinos is the relic neutrino background of the universe \cite{PhysRevLett.115.091301}, but a conceptual basis for observing it has yet to be determined. 
Recently, the effective potential between aggregate matter mediated by the exchange of two non-relativistic virtual neutrinos has been derived \cite{PhysRevD.101.093004} near its range with all ingredients for the neutrino properties included - it is consistent with and predicted by the Standard Model of Particle Physics. 
This is a coherent $\Delta L = 0$ interaction of matter with a weak charge proportional to the number of constituents, \textit{implying a violation of the equivalence principle} with different weight of electrons, protons, and neutrons with respect to the mass of matter. 
As the Compton wavelength of neutrinos of mass $0.1$ eV or less is beyond $\SI{1}{\micro\meter}$, this regime gives the range of the interaction.
At these distances, the exchanged neutrinos are non-relativistic and so this weak potential is sensitive to the absolute neutrino mass and the Dirac/Majorana distinction, see Fig.~\ref{fig:1}. 
Contrary to the concept of $\Delta L = 2$ transitions existing for Majorana neutrinos only, the $\Delta L = 0$ interaction discussed here exists for both Dirac and Majorana neutrinos, and are distinguishable by mass. 
Additionally, and unlike the case of $\Delta L = 2$ searches, the $\Delta L = 0$ interaction is not heavily dependent on the Pontecorvo-Maki-Nakagawa-Sakata (PNMS) mixing parameters or the ordering of neutrino masses.
With this interaction, it now becomes conceptually feasible to determine both the mass and nature of neutrinos in a single experiment, which sets it apart from alternative approaches reliant  on beta decay kinematics and neutrinoless
double beta decay searches separately.   
\par In this article, we explore the prospect of observing this weak, repulsive potential generated by the non-relativistic, virtual exchange of two neutrinos - the so-called $2\nu$ potential. 
For this goal we maximize the $2\nu$ potential on an atomic probe $P$, from geometric considerations, for a distributed and extended source material $S$; we are led to a cylindrical source with an axial bore. 
In Section II we discuss the distance behavior of the potential and its weak charges for $P$ and $S$, in the limit of massless neutrinos, for finding the optimal thickness of $S$ where the effective potential saturates. 
Section III extends this analysis to massive neutrinos, them being either Dirac or Majorana particles, which changes the distance scaling and the weak charges of the six pairs of neutrinos with definite mass and flavor mixing. 
The dependence of the resulting potential between the extended source $S$ and the atomic probe $P$ on the absolute mass and the nature of the mediating neutrinos is a demonstration of the conceptual basis of this work.  
Given the range of microns for the potential and the corresponding scenario where these sensitivities are apparent, in Section IV we consider how previous low energy atomic and neutron experiments, in the micron regime from a surface, place bounds on this potential. 
This has been considered previously \cite{Dinh2013,Stadnik2018,Ghosh2020,DeMille2017,PhysRevLett.129.239901}, but new experiments and associated theoretical work offer significant promise for improvement. 
In {Section V} we address dominating background potentials from gravitational to Casimir-Polder effects: utilization of different strategies for the construction of the source material and spatial scaling allow the $2\nu$ potential to be isolated. 
Section VI follows by proposing a schematic experiment, with the expectation that  future experimental progress could open the window to a feasible tabletop experiment to explore the potential generated by the $2\nu$ exchange interaction, and so the absolute mass and the Dirac or Majorana particulate nature of the neutrino. 
We then present the conclusions and outlook projected by this work.
%
%
%
%
%
%
%
%
\section{THE $2\nu$ POTENTIAL AND THICKNESS OF ITS SOURCE}
\begin{figure*}
\centering
\includegraphics[width=\linewidth]{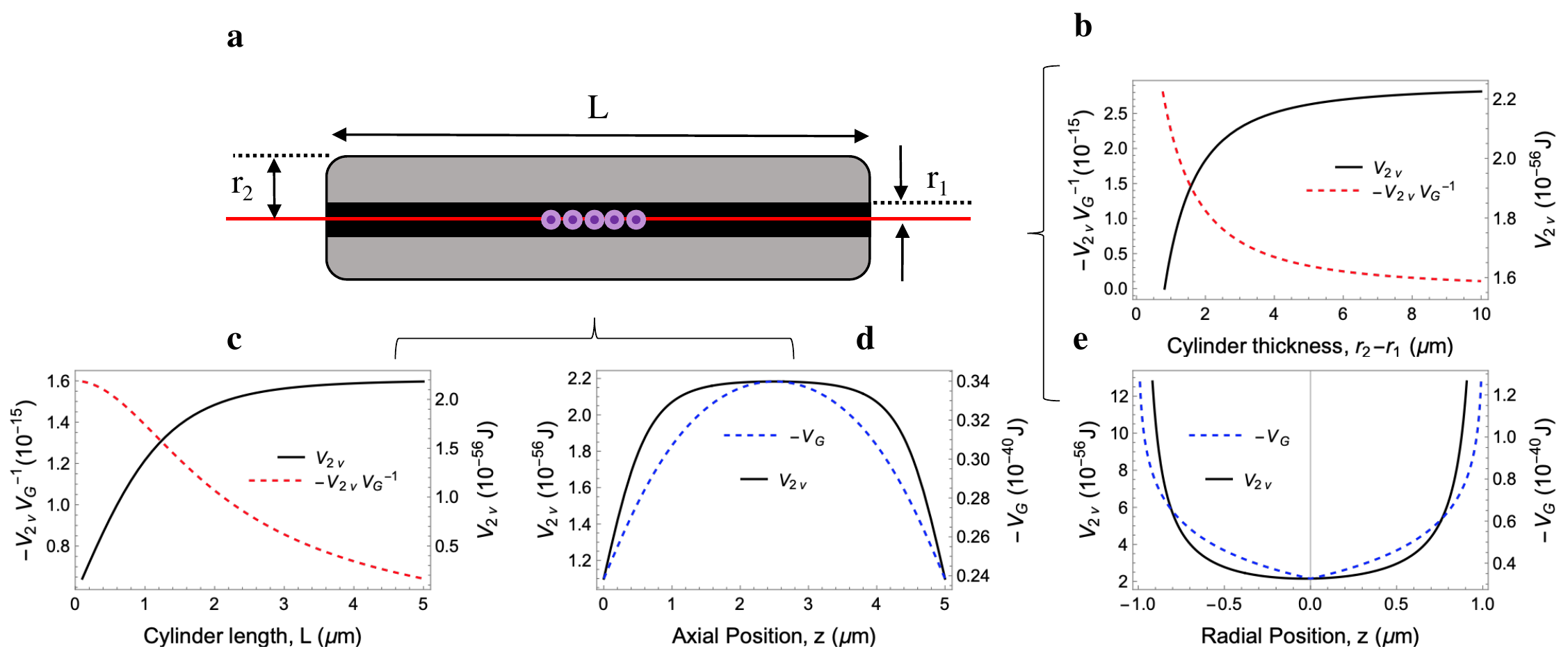}
\caption{
Source geometry optimization for generating a 2$\nu$ exchange potential.
\textbf{a}: Cylinder with a bore hole, definition of variables, proposed location of probes (purple) along centerline (red) of a cylindrical source.
\textbf{b}: Ratio between the 2$\nu$ exchange and the gravitational potential as function of the cylinder thickness and \textbf{c}: length. 
In both cases we show (solid lines, right y-axis) the dependency of 2$\nu$ potential compared to the ratio of the potential with gravity from the source (dashed lines, left y-axis).
Note that the 2$\nu$ potential saturates around $\SI{4}{\micro\meter}$.
We have optimized the potentials along the center line of a cylinder with a thickness of $\SI{4}{\micro\meter}$.
Note the different scales of the potentials.  
\textbf{d}: As a function of the position along the axis, we show the 2$\nu$ potential (solid line) is more flat compared to gravity (dashed line) and
\textbf{e}: as a function of the radial position in the center of the cylinder.
Note the gravitational instability at the center. 
\label{fig:2}} 
\end{figure*}
\par The 2$\nu$ potential has a range of microns \cite{PhysRevD.101.093004}, see Fig.~\ref{fig:1}, and is built from the t-channel absorptive part of the scattering amplitude while remaining dependent upon shell neutrino properties. 
Neutral aggregate matter has a coherent weak charge proportional to the number of constituents: $Z$ electrons and protons and $N$ neutrons.
The properties of the potential depend upon the pair of exchanged neutrinos of definite mass. 
Due to the mixing with electron neutrinos, electrons are affected by charged current weak interactions in addition to the neutral current weak interactions with all neutrino flavors.
Protons and neutrons have universal diagonal neutral current interactions with all neutrino species. 
Neutral current couplings of electrons and protons are of opposite sign, simplifying the calculation of the Hermitian matrix $Q^{ij}_{W} = 2 Z U_{ei} U^{*}_{ej} - N \delta_{ij}$ of the six coherent charges of matter for each $(i, j)$ pair of definite-mass neutrinos, where $U$ is the PMNS mixing matrix. 
In the limit of massless neutrinos, the spatial dependence of the potential between the source and the probe becomes a unique inverse power law. 
It is in this way we may construct the sum of the products of the six charges for the source and the probe, leading to an inoperative mixing and (see Eq. (7) of Ref. [9]) having a sum of products of diagonal flavor charges of the matter $Q^{e}_{W} = 2Z - N$ for electron neutrinos and $Q^{\mu, \tau}_W = -N$ for muon and tau neutrinos. 
$Q^{e}_{W}$ develops due to charged current weak interaction of the $Z$ electrons and the neutral current interaction of the $N$ neutrons, whereas $ Q^{\mu, \tau}_{W}$ is due to the neutral current of the $N$ neutrons. 
Given this, the $2\nu$ potential for massless neutrinos can be written as:
%
%
\begin{equation}
    V_{SP}(R) = \frac{V^0_{SP}}{R^5}
\end{equation}
where $R$ is the distance between the source $S$ and the probe $P$ and
\begin{equation}
   V^0_{SP} = ((2Z_S-N_S)(2Z_{P}-N_{P})+2N_{S} N_{P})\frac{G_F^2 (\hbar c)^5}{16\pi^3}
\end{equation}
where $G_{F}$ is the Fermi coupling constant of weak interaction, with $Z_{S,P}$($N_{S,P}$) being the atomic (neutron) number of the two atomic species. 
In what follows we take a single atomic probe and a distributed source of the potential, with a number density $\rho$, and in doing so move from atom-atom to atom-volume interactions, with the volume limited by two cylindrical surfaces of radii $r_{1}$ and $r_{2}$, see Fig.~\ref{fig:2}\textbf{a}.
We have considered several possible geometries, such as planar, spherical, needle-like, and pyramidal, but we find that a cylindrical tube is a practical compromise for optimizing the field.
In cylindrical coordinates $(r, \phi, z)$, the potential induced on a single confined particle at the center,taken as the origin $r = z = 0$, from a source volume element of particles with the same $(Z_{S}, N_{S})$ is given by
\begin{equation}
d V_{SP}(R) = \frac{\rho V^0_{SP} r}{R^5} dr~d\phi~dz
\end{equation}
where $R^{2}=r^{2}+z^{2}$, (See Appendix A for more details).
The potential induced by the cylinder is only $3/2$ times smaller than equivalent spherical geometries, but such a geometry lends itself significant ease of use through analytical solutions to the potential and experiment design.
For reasons to be explained shortly, we choose $_{32}^{74}\text{Ge}$ as the source and $_{70}^{174}\text{Yb}$ as the atomic probe for the 2$\nu$ potential (see Appendix B), the atoms used in Fig.~\ref{fig:1} and the rest of the calculations considered here.
\par We use the computed  potentials to optimize the geometry of the source: inner radius $r_1$, outer radius $r_2$, and length $L$, see Fig.~\ref{fig:2}\textbf{a}. 
We fix the inner radius $r_1 = \SI{1}{\micro\meter}$, where the interaction departs from the massless neutrino case, see Fig. \ref{fig:1}, and study the behavior of the weak potential benchmarked against the gravitational potential from the source as function of $r_2-r_1$, Fig.~\ref{fig:2}\textbf{b}, and $L$,  Fig.~\ref{fig:2}\textbf{c}.
The 2$\nu$ potential decreases with distance faster than the gravitational potential; consequently, the ratio between them for the interaction between the source and the probe decreases. 
The potential increases with the volume of the source until it saturates at about $\SI{4}{\micro\meter}$ due to its fast decrease with distance compared to the increasing volume, proportional to $L r_2^2$.
This saturation behavior is in stark contrast with that for the gravitational potential, and so fixes the appropriate choice of $r_{2}$ of the cylinder.
The scale differs between Fig.~\ref{fig:2}\textbf{b} and \textbf{c}, due to this differing dependency on the volume of the source in $L$ and $r_2$.
We conclude that a finite cylinder thickness contributes to an enhancement of the 2$\nu$ exchange potential up to values reached at about $\SI{4}{\micro\meter}$.                 
At these distances, the effect of the neutrino mass and its Dirac/Majorana nature will emerge\cite{PhysRevD.101.093004}, as seen in Fig.~\ref{fig:1}.
\par Along the axis of the source, see the solid red line of Fig.\ref{fig:2}\textbf{a}, the 2$\nu$ potential has a weaker curvature than that of gravity, Fig.~\ref{fig:2}\textbf{d}.
When considered radially, Fig.~\ref{fig:2}\textbf{e}, the weak curvature is evident as well, with the addition of a gravitational instability (a discontinuity) at the center.
This could prove useful, as a way of distinguishing the 2$\nu$ potential from the gravity of the source.
\section{SENSITIVITY TO THE NEUTRINO MASS AND NATURE}
\begin{figure}
\centering
\includegraphics[width=.9\linewidth]{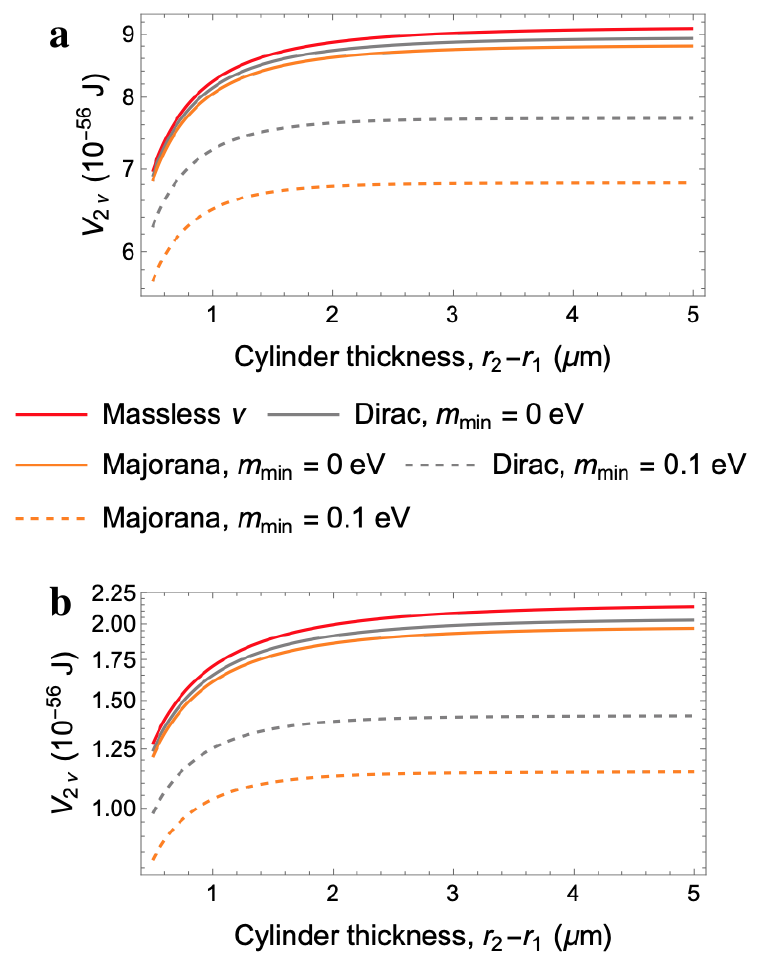}
\caption{ Integrated potential at the centre of a cylinder of inner radius \textbf{a}: $\SI{0.5}{\micro\meter}$ and \textbf{b}: $\SI{1}{\micro\meter}$ with a $\SI{5}{\micro\meter}$ length, shown as function of the cylinder thickness, for five cases.  
\label{fig:3}}
\end{figure}
\par The dependency with any given mass of the exchanged neutrino pair invalidates the $R^{-5}$ potential, implying the existence of a symmetric $3\times3$ weak charge matrix with known flavor mixing for both the source and the probe. 
These masses are determined by neutrino oscillations in terms of an absolute scale, like the mass of the lightest neutrino. 
With non-vanishing neutrino masses, the spatial dependency of the potential beyond $\SI{1}{\micro\meter}$ is short range. 
This virtual non-relativistic neutrino pair is a $p$-wave for Majorana neutrinos, contrary to the Dirac case. 
Nevertheless, for each of the cases shown in Fig.~\ref{fig:1}, the atom-atom potential between $\SI{1}{\micro\meter}$ and $\SI{10}{\micro\meter}$ is approximated by 
\begin{equation} 
 V^i_{2\nu} = \sum_{n=1}^5 \frac{\alpha^i_n}{R^{n}}, 
\end{equation}
where the superscript $i$ is each case (see Appendix C for details).
The five cases considered are that of all massless neutrinos, vanishing lightest mass for Dirac or Majorana, and the lightest neutrino being of mass $0.1$ eV for Dirac or Majorana. 
The atom-atom potential is incorporated into the number density-dependent potential for the extended cylindrical source. 
The integrated potentials are shown in Fig.~\ref{fig:3}\textbf{a} and \textbf{b} as a function of the cylinder thickness, for an inner radius $r_1 =\SI{0.5}{\micro\meter}$ and for $r_1 =\SI{1}{\micro\meter}$.
The 2$\nu$ potential field strength exhibits a strong dependency on nature of the neutrino.
\par The results obtained for the atom probe-cylindrical material source at micron distances show the expected behavior of the decrease of the 2$\nu$ potential with increasing inner radius of the cylinder and the increase of the sensitivity to the neutrino absolute mass and its nature. 
Conceptually, there exists the possibility to explore these properties using the approach shown. 
It is a novel method, entirely independent of the $\Delta L = 2$ experiments that allow for Majorana neutrinos only, and it moves the burden of proof to a fundamentally lower energy scale where the neutrino mass term emerges. 
\section{Prior experiments}
\par Is there a plausible detection scenario? 
The $2\nu$ potential is minute, at micron distances being orders of magnitude weaker ($10^{-15}$) than that of gravity between the probe and the source, with the magnitude being $10^{-56}$ J at $\SI{1}{\micro\meter}$ - see  Fig.~\ref{fig:3}. 
This is well below the direct detection limit of existing low energy particle and atomic physics experiments. 
The atom-volume interaction, along with improvements from considering larger weak charge probes (see Appendices), are enhancements that make it worthwhile to consider previous experiments while also provoking thought experiments that could lead to a future bespoke low energy experiment. 
We start by considering a selection of prior experiments utilizing probe particles near surfaces.
\par A series of experiments in the micron regime have been performed in atomic and neutron physics, often studying Casimir-Polder interactions \cite{Laliotis2021} and limits on new Yukawa-type short range forces \cite{193548}. 
The geometries employed in these experiments involved planar sources, which are considered unfavorable despite their analytical solubility. 
The calculation of the Casimir force between plates for neutrinos with two specific masses in the loop has been previously conducted \cite{Costantino_2020}. 
However, resolution of the $2\nu$ potential in these experiments would necessitate a sensitivity of approximately $10^{-60}$ J at a distance of $\SI{1}{\micro\meter}$.
We focus on experiments in the range $\SI{0.5}{\micro\meter}$ to $\SI{10}{\micro\meter}$ as a region of interest and convert results from prior works to approximate equivalent potentials for the continued illustration of our point:
An atomic beam of neutral $^{23}$Na passing between two parallel plates with a variable spacing between $\SI{0.7}{\micro\meter}$ and $\SI{7}{\micro\meter}$ resolved the Casimir-Polder potential to $1.7 \times 10^{-28}$ J - and was found to be in complete agreement with theory \cite{PhysRevLett.70.560}. 
Modern ultracold bouncing neutron experiments \cite{Jenke2011} demonstrating gravitational resonance spectroscopy resolved energy differences of $3 \times 10^{-33}$ J in the tens of $\SI{}{\micro\meter}$ regime. 
Neutrons are a notable option as a probe, despite a low weak charge, as they are essentially undisturbed by residual electromagnetic forces, such as the Casimir-Polder interaction, given that its electric polarizability is small. 
Measurements of the change in induced oscillation frequencies in magnetically trapped Bose-Einstein condensates of alkali atoms $\SI{6}{\micro\meter}$ to $\SI{9}{\micro\meter}$ above a surface \cite{PhysRevA.72.033610} realized a sensitivity nearing $5 \times 10^{-37}$ J. 
Atom interferometry on a chip \cite{Alauze2018} observed $3 \times 10^{-38}$ J about $\SI{10}{\micro\meter}$ above a surface, with one hour of integration and with $\SI{3}{\micro\meter}$ resolution. 
State-of-the-art atom interferometry, run in free space and as a differential experiment \cite{Parker2018}, is sensitive enough to be only one order of magnitude away from observing QED corrections, a promising lead. 
An atomic clock trapped on a chip \cite{PhysRevA.92.012106} provides bounds in the region $\SI{0.6}{\micro\meter}$ to $\SI{10}{\micro\meter}$ to $7 \times 10^{-40}$ J. 
All of these experiments are critically constrained by sub-optimal planar geometry, source/probe composition, and the effects of surface physics which was often the point of interest in these works, and so demonstrate a lack of sensitivity by orders of magnitude by comparison to cylindrical or spherical geometries.
There are no experiments that place favorable bounds from low energy on the $2\nu$ potential. 
\section{Theoretical Backgrounds}
\par Further and contrary to mass density and number density dependent potentials which govern, respectively, the gravitational and weak interaction, the Casimir-Polder interaction is controlled by the electronic structure of the material source through its electric polarizability \cite{PhysRevD.101.093004}.
Most prior experiments in the micron regime occurred near metallic surfaces (gold being a common surface substrate, over silicon). 
Choice of a pure non-polar covalent material in a high quality detector grade crystal, like diamond, silicon, or germanium, allows for a suppression of Casimir-Polder effects. 
For our proposed experiment, we choose germanium versus similar materials based on it having a high weak charge for each atom and given that it can be grown with a low impurity level - this can be accomplished by re-growing several times with Czochralski's method \cite{Czochralski_1918,Singh2012}. 
The isotopic disorder of germanium crystals can be taken into account by using the average weak charge and the number density of atoms - this approximation demonstrates a near zero electronic polarizability of the entire source in the region of interest.
The absence of electric polarizability for the material source, as expected from its tetrahedral crystalline structure, would be confirmed with Raman and other infrared spectroscopic techniques.
Another option consists of materials constructed from dielectric nanotubes, expected to have
low values of the transverse component of the electric polarizability\cite{Lu2007,Lan2009}.
We summarize values of the various potentials for combinations of probe and source elements in Table~\ref{Tab:AtomicOptions}, found in Appendix B.
\begin{figure}
\centering
\includegraphics[width=0.8\linewidth]{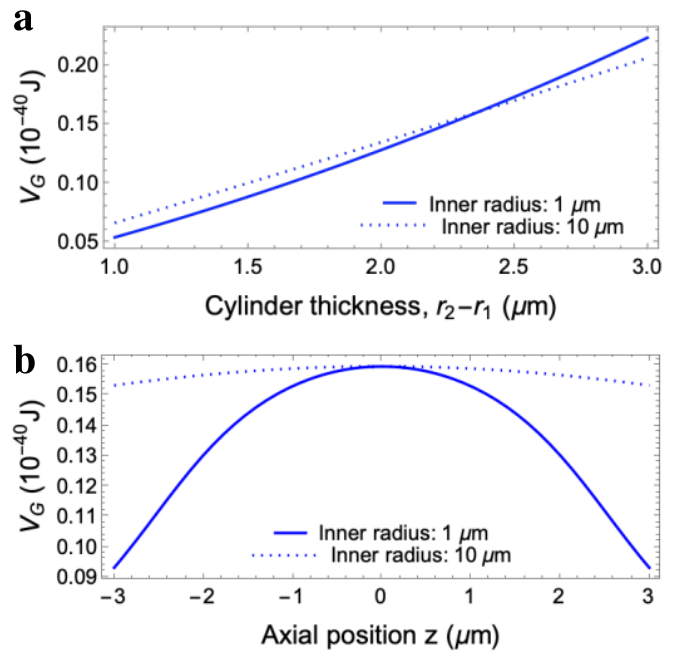}
\caption{Demonstration of the equalization of the gravitational potential at the center of two source cylinders of inner radius \textbf{a} $\SI{1}{\micro\meter}$ (solid line) and \textbf{b} $\SI{10}{\micro\meter}$ (dashed line) and $\SI{5}{\micro\meter}$ length. 
The potentials are computed for different cylinder thickness, equal for both geometries, in \textbf{a}. 
The potential equalization occurs for a common thickness of $\SI{2.362}{\micro\meter}$. 
In \textbf{b}, the potential along the cylinder axis for the magic compensation value of $\SI{2.362}{\micro\meter}$. 
\label{fig:4}} 
\end{figure}
\par The gravitational interaction cannot be suppressed. 
However, in an interferometric experiment we can try to suppress the effect by compensation within the arms of the interferometer without affecting the
$2\nu$ potential; the differing spatial dependence of the two interactions provides a conceptual solution. 
By example, we take two cylinders of the same density and length, one with an inner radius of $\SI{10}{\micro\meter}$ and another with $\SI{1}{\micro\meter}$.
The weak potential is negligible for the large radius cylinder, but the gravitational potential at the center can be made equal to that of the smaller radius cylinder by compensating the longer distance with a larger mass at a magic width, common to the two cylinders - see Fig.~\ref{fig:4}. 
This allows for a pathway to disentangle local source gravity from the $2\nu$ potential, in addition to using a combination of different properties for pattern generation \cite{Hudson2014}, like spatial dependence and different atom probes looking for a violation of the equivalence principle.
\section{A schematic experiment}
\par
\begin{figure}[!ht]
\centering
\includegraphics[width=\linewidth]{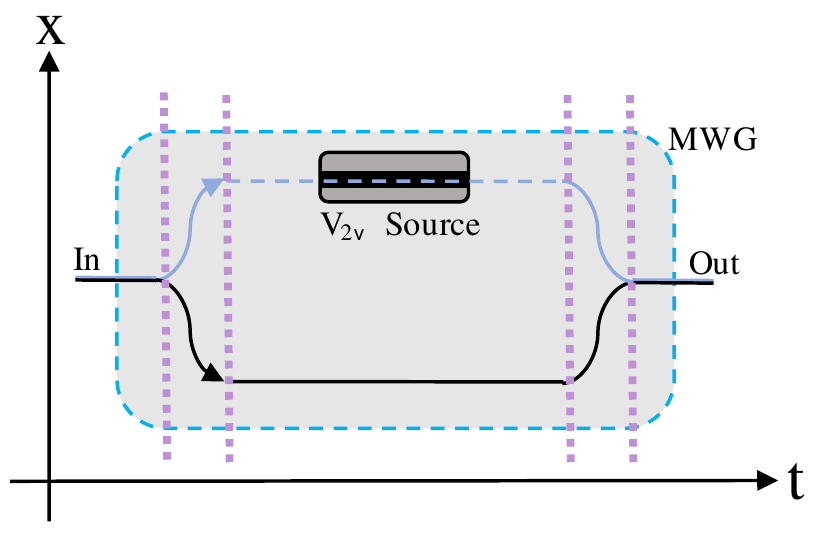}
\caption{Proposed spacetime diagram of a schematic atom potentiometer. 
Central trajectories for a hybrid atom interferometer, with one arm through a $V_{2\nu}$ source (periwinkle).  in the laboratory frame where we propose a Ramsey-Bord\'{e} sequence (purple dashed lines are the interferometer pulses) in a magnetic waveguide (MWG). 
The waveguide functions as a holding potential (grey region), to increase the interferometer time inside the $V_{2\nu}$ source.
The clock path alternatives are parallel and are initialized simultaneously. 
The signal of interest is a frequency shift between the two arms, read out as a phase difference depending upon the source mass geometry, time held in the source, and systematics.
\label{fig:5}} 
\end{figure}

 With this last consideration, we turn our attention to recent developments in atomic physics, especially techniques from metrology, as a starting point for a possible experiment. 
The physics of the $2\nu$ potential gives a series of additional directives for the experiment:  heavy, neutral atoms, and so with an appreciable weak charge, as the probe and a dense covalent source material, so as to suppress  electromagnetic effects and that can be precisely constructed via crystal growth.
Given this, we chose $^{174}$Yb as the probe and $^{74}$Ge as the source for all the work shown in this article - see Table~\ref{Tab:AtomicOptions} (Appendix B) for different combinations of probe and source species.
Such an experiment as this is sensitive to the positioning of the probe with respect to the axis of the cylindrical source, as is demonstrable from Fig.~\ref{fig:2}; utilizing an ultracold quantum gas in a magnetic waveguide is a plausible probe scenario, to keep the experiment in the interaction volume at the point of highest symmetry.  
An advantage of ultracold gases is the ability to generate unique states of matter based on dimensionality, scattering length, and temperature; a preparation as one-dimensional hard bosons \cite{Kinoshita2004-fi} would be ideal, with minimal atom-atom (probe-probe) interactions. 
We envision a chain of probes numbering $N_{\text{atoms}} \leq 100$ atoms in the two arms of an atomic clock interferometer \cite{Loriani2019,PhysRevX.10.021014}, see Fig.~\ref{fig:5}. 
The two clocks, initialized at the same time and running parallel to each other, are held in free space above a magnetic waveguide, and/or within a series of cylindrical masses like that of Fig.~\ref{fig:2} (the main experiment) and modified like that of Fig.~\ref{fig:4} (systematics, source gravity effects). 
These clocks are functioning as local atomic potentiometers, held for a long time between the sets of pulses of the interferometer.  
Such an experiment would be enhanced with detection utilizing single-atom methods \cite{PhysRevX.9.041052, Bakr_2009} and spin-squeezing \cite{greve22}. 
\par Recent work on quantum-clock interferometry \cite{Loriani2019,PhysRevX.10.021014,Bothwell_2022} for studying gravitational redshift gives a relevant and recent formalism to consider the interference of atomic clocks as an atom poteniometer; we show, in a schematic experiment like Fig.~\ref{fig:5}, that a differential phase generated by the 2$\nu$ potential near a matter source in one arm of the clock interferometer exists. 
We consider the interference pattern generated by the two arms with different
environments, be it either two different matter sources or a single matter source compared to free space propagation one has a probability $\mathcal{P}$ at recombination: 
\begin{equation}
\begin{split}
\mathcal{P}&:=\braket{\Psi\vert\Psi}=\lvert\psi(x,t)\rvert^{2}\\
&\;=\frac{1}{2}[1+C\cos{(\Delta\phi)}]
\end{split}
\end{equation} 
where the interferometric contrast $C=\lvert\braket{\Phi_{\text{source}}\vert\Phi_{\text{free}}}\rvert$, is given by the overlap of the wave functions for the two alternative paths in space-time, through the source and free space in Fig~\ref{fig:5}.
Fundamentally, the differential phase has terms that can be grouped as 
\begin{equation}
\Delta\phi=\delta\phi_{\tau}+\delta\phi_{\text{las}}+\delta\phi_{\text{pot}},
\end{equation}
where $\delta\phi_{\tau}$ is the shift from finite initialization time difference (governed by the spatial distance between the arms and the speed of light) and $\delta\phi_{\text{las}}$ from the light pulses. 
This experiment is kept along the same geodesic, such that any tilt from perpendicular to Earth's gravity is minimized - this suppresses a series of terms that would contribute to $\delta\phi_{\text{pot}}$, the term of interest. 
This term contains phase shifts from residual Earth gravity, gravity gradients, and rotations, source gravity, and Casimir-Polder effects. 
There is a residual phase shift $\phi_{\nu\nu}$ from the two neutrino exchange potential contained within this term, $\delta\phi_{\text{pot}}\ni\phi_{\nu\nu}$, and it is given by
\begin{equation}
\phi_{\nu\nu}(r,t)\simeq\frac{t}{\hbar}V^i_{2\nu}(r),
\end{equation}
where $t$ is the time spent within the field interacting. 
Our point is now ready for a demonstration: with our choice of probe and matter source of $\SI{1}{\micro\meter}$ and for $10$ s, we find a phase shift $\phi_{\nu\nu}\sim 10^{-22}$~rad exists in principle, but this is inaccessible. 
Further enhancement of the potential is an unlikely development; we look toward improvements in atomic and metrological technique and extended time experiments. 
\par Designing and constructing an experiment to start with exploring source gravity physics at micron distances should be made.
Moving forward is to involve new results from atomtronics \cite{kim2022}, entanglement protocols \cite{Madjarov_2020,greve22}, cryogenic cooling of the source for improved vacuum and reduced blackbody and Casimir effects, as well as high fidelity single atom detection \cite{PhysRevLett.120.103201,PhysRevX.9.041052}, to constitute a new state-of-the-art experiment in an attempt to explore the two neutrino exchange potential from low energy with precision atomic physics.
\section{Concluding remarks}
\par We have explored the prospect of observing the weak, repulsive potential generated by the non-relativistic, virtual exchange of two neutrinos at micron distances.
Within this range, the size of the potential is dependent upon the absolute mass and the fundamental particulate nature of the neutrino. 
We have shown an optimization of this potential on an atomic probe from geometric considerations, generating a matter source with the criterion that the potential saturates. 
This shows a 15 order of magnitude improvement over the atom-atom interaction previously calculated. 
We consider prior low energy atomic and neutron experiments in the micron regime and find that planar geometries and insufficient precision place no meaningful bounds on such a potential.  
We propose to suppress residual Casimir-Polder interactions by the use of a purely covalent material like germanium. 
A schematic experiment is outlined to begin exploring this potential, starting with source gravtity at micron distances, and that could lead to a low energy bounding of the $2\nu$ potential probes.
The gravity effect can be compensated, using an appropriate matter sources in different arms as well as different probes - this differential experiment is necessary to elucidate the violation of the equivalence principle by this weak potential.
This last comparison makes apparent the violation of the equivalence principle implied by the weak charge of aggregate matter compared to its mass.
Despite this new understanding, distinct non-vanishing results of the potential remains out of reach for existing state-of-the-art low energy precision measurements; future experimental progress could open this window with a low energy precision measurement tabletop experiment.
\par Here we show a new understanding of the accessibility of atomic experiments to the open problems of neutrino mass and nature. 
While the expected results for the detection of the weak potential between aggregate matter at micron distances remain out of reach, the complementarity of our concept, based on quantum physics, make this approach worthwhile to pursue. 
Contrary to the search of $\Delta L = 2$ processes for Majorana neutrinos, this
$\Delta L = 0$ effect would provide distinct results between Dirac and Majorana mass terms.  
Attaining such sensitivity to the weak potential could elucidate both the absolute neutrino mass and its nature without the need for additional assumptions, all within a single experiment. 

\begin{acknowledgments}
We would like to thank Andres Cantarero and Pablo Rodríguez for enlightening discussios on the properties of dielectric materials. 
JB has been supported by MICINN/AEI, PID2020-113334GB-I00/ AEI / 10.13039/501100011033, and CIPROM/2021/054 (Generalitat Valenciana).
FS thanks the Swiss National Foundation under the grant No. 200020\_204609.
All authors warmly thank the Boninchi Foundation for their generous support for this project. 
\end{acknowledgments}
\section*{Data Availability}
The data presented and analyzed in this study are available from the corresponding author upon reasonable request. 
\section*{Conflict of interest}
The authors have no conflicts to disclose.
\newpage
\bibliography{sn-bibliography}
\newpage
%
%
%
%
%
%
%
%
%
%
%
\newpage
%
\appendix
\section{Potential created by a source distributed in a cylinder}
\label{App:cylinder}
\par Let's consider the case of a long tube with inner radius $R_{\text{in}}$ and wall thickness $T$. 
In an infinite long tube, 
\begin{equation}
\begin{aligned}
V_{2\nu} &= \frac{4\pi \rho V^0_{SP}}{3}\left( \frac{1}{R_{\text{in}}^2}-\frac{1}{(R_{\text{in}}+T)^2}\right),
\end{aligned}
\end{equation}
We have computed that the cylinder potential is four times larger than the one provided by an infinite plane and 2/3 times the one at the center of a source sphere with the same material thickness.
The $2\nu$ potential as function of the radial position of the probe inside the cylinder ($r$) is given by:
\begin{equation}
\begin{aligned}
 V_{2\nu}(r) =& \frac{4\pi \rho V^0_{SP}}{3}\\
 &\left( \frac{1}{R_{\text{in}}^2-r^2}-\frac{1}{(R_{\text{in}}+T)^2-r^2}\right)
\end{aligned}
\end{equation}
The potential is rather flat around $r = 0$ and changes rapidly when $r \approx R_{\text{in}}$ (i.e. the probe gets closer to the source walls). At the center of the tube ($r=0$) we recover the previous result.
\par The same calculation applied to a gravitational potential predicts an infinite potential associated to the infinite mass of the cylinder. A more realistic calculation considers a finite length of the source ($H$). The potential as function of the location of the probe inside the cylinder can be also computed analytically for different probe locations,  
\begin{equation}
\begin{aligned}
V_{2\nu}(r,z) = \rho V^{0}_{SP} \pi  \left(\frac{1}{R_{\text{in}}^{2}-r^{2}}+\frac{1}{r^{2}-(R_{\text{in}}+T)^{2}}\right)\\
\bigg\lvert\frac{3 \left((R_{\text{in}}-r)^{2}+(H-z)^{2}\right) (z-H)-(z-H)^{3}}{3 \left((R_{\text{in}}-r)^{2}+(H-z)^{2}\right)^{3/2}}\\
+\frac{z^3-3 z \left((R_{\text{in}}-r)^2+z^2\right)}{3 \left((R_{\text{in}}-r)^2+z^2\right)^{3/2}}\bigg\rvert,
\end{aligned}
\end{equation}
where $z$ is the coordinate along the cylinder axis, $r$ the radial position with respect the cylinder axis, $R$ is the inner cylinder radius, $T$ the cylinder wall thickness and $H$ the cylinder length.  
\par On the contrary, the gravitational potential induced by the source in the probe shows different dependency with the tube thickness ($W$). We have to consider a finite tube of length ($H$):
\begin{equation}
\begin{aligned}
V_G(r,z) =& -2 \pi M_{P} \rho G \bigg( 2 R_{\text{in}} T + \frac{T^{2}}{2}\bigg) \\
&\bigg( \tanh ^{-1}\left(\frac{z-T}{\sqrt{(H-z)^{2}+(R_{\text{in}}-r)^{2}}}\right) \\
& -\tanh ^{-1}\left(\frac{z}{\sqrt{(R_{\text{in}}-r)^{2}+z^{2}}}\right)\bigg)
\end{aligned}
\end{equation}
where $M_P$ is the atomic mass of the probe, $r$ is the radial position with respect the cylinder axis and $z$ the coordinate along the cylinder. The potential dependencies with the cylinder radius and length permits the operation of two cylinders, one with vanishing 2$\nu$ potential and the other with a sizeable potential but sharing the same gravitational potential. 
\section{Influence of the atomic species}
\label{App:AtomSpecie}
The optimal source potential is the one that optimises $V^0_{SP}\rho$. 
We have computed the quantity for several source and probe species.
The relative strength to that of the gravitational force goes as $V^0_{SP}$. 
Values are tabulated in Table~\ref{Tab:AtomicOptions}. 
We can gain a factor of $\sim$10 in atom-atom potential strength ($V^0_{SP}$) from carbon to tungsten, and a factor of 5.5 when considering the density of the source material ($V^0_{SP}\rho$).  
The lost of strength by changing the $_{74}^{184}\text{W}$ source by  $_{32}^{74}\text{Ge}$ is around 3.6. 
We consider other aspects related to the control of the Casimir-Polder potential in the selection of the source material. 
We have also computed the same potential for a lighter probe atom ($^{40}_{19}\text{K}$), see Table~\ref{Tab:AtomicOptions}. 
The $^{40}_{19}\text{K}$ potential strength  versus the one with $_{70}^{174}\text{Yb}$ varies from 4.5 to 4.9 times depending on the source atom specie. 
The same ratio is obtained for $V_{SP}\rho$ since the density depends only on the source component. 
This results shows, as expected, that the optimal configuration is the one with the largest density in the source and largest number of nucleons in the source and probe species. The case of the carbon source is special since it is possible to achieve high density material with respect the nominal atomic mass using graphite as a source material. The $V_{SP}\rho$ strength has a similar magnitude to the one of  $_{32}^{74}\text{Ge}$  despite a large difference in the atom-atom potential ($V_{SP}$) between the two source atomic species. 
\begin{table*}
\begin{center}
\caption{ Values of $V^0_{SP}$ and $V^0_{SP}\rho$ for different combinations of atoms. Density for $^{12}_{6}\text{C}$ is taken for graphite.}
\begin{tabular}{|c |c| c | c | c |} 

 \hline
Probe & Source & $\rho$ $(\SI{}{atoms}/\SI{}{\micro\meter}^3)$ & $V^0_{SP}$ $(\SI{}{\joule}\,\SI{}{\micro\meter}^5)$ & $V^0_{SP}\rho$ $(\SI{}{atoms}\, \SI{}{\joule}\, \SI{}{\micro\meter}^2)$\\ 
 \hline
 \hline
$_{40}^{174}\text{Yb}$ & $_{6}^{12}\text{C}$ & $1.13 \times 10^{11}$ & $3.25 \times 10^{-68}$ &  $3.68\times10^{-57}$ \\ 
$_{40}^{174}\text{Yb}$ & $_{32}^{74}\text{Ge}$ & $4.41 \times 10^{10}$ & $1.24 \times 10^{-67}$ &  $5.47\times10^{-57}$ \\ 
$_{40}^{174}\text{Yb}$ & $_{74}^{184}\text{W}$ & $6.31 \times 10^{10}$ & $3.15 \times 10^{-67}$ &  $1.99\times10^{-56}$ \\ 
\hline
\hline
$_{19}^{40}\text{K}$ & $_{6}^{12}\text{C}$ & $1.13 \times 10^{11}$ & $6.56\times10^{-69}$ &  $7.44 \times 10^{-58}$ \\ 
$_{19}^{40}\text{K}$ & $_{32}^{74}\text{Ge}$ & $4.41 \times 10^{10}$ & $2.78\times 10^{-68}$ &  $1.23\times 10^{-57}$ \\ 
$_{19}^{40}\text{K}$ & $_{74}^{184}\text{W}$ & $6.31 \times 10^{10}$ & $6.85 \times 10^{-68}$ &  $4.32 \times 10^{-57}$ \\  
 \hline
\end{tabular}
\label{Tab:AtomicOptions}
\end{center}
\end{table*}
\section{Atom-Atom $2\nu$ potential expansion in terms of $1/R^{n}$}
To simplify numerical integrals, we approximate the atom-atom potential between 1~$\mu m$ and 10~$\mu m$ by an expansion in terms of $1/R^n$, where n is a positive integer. 
The integral are limited to values between these two limits to ensure the validity of the expansion. 
This expansion is given by:
\begin{equation} 
 V^i_{2\nu} = \sum_{n=1}^5 \frac{\alpha^i_n}{R^n}, 
\end{equation}
Here, the superscript i represents each of the cases that have been analyzed. The values of $\alpha_i^n$ for the potential induced between $_{70}^{174}\text{Yb}$ and $^{74}\text{Ge}_{32}$ atoms can be found in Table~\ref{Tab:ExpansionParameters}. 
The reference case for massless neutrinos, with a nominal dependency of $1/R^5$, is also shown.
\begin{table*}
\begin{center}
\caption{Expansion coefficients $\alpha_i^n$ representing the $1/R^n$ potential between ${74}^{32}\text{Ge}$ and ${70}^{174}\text{Yb}$ atoms.}
\begin{tabular}{|l |c| c | c | c |c|} 
 \hline
 $m_{\text{min}}$  & $\alpha_1 (J\, $\SI{}{\micro\meter}$)$ & $\alpha_2 (J\, $\SI{}{\micro\meter}$^2)$ & $\alpha_3 (J\, $\SI{}{\micro\meter}$^3)$ & $\alpha_4(J\, $\SI{}{\micro\meter}$^4)$ &$\alpha_5(J\, $\SI{}{\micro\meter}$^5)$\\
 \hline 
  \multicolumn{6}{c}{Massless} \\
 \hline
   0~eV & - & - & - & - & $1.26 \times 10^{-67}$\\
\hline 
\multicolumn{6}{c}{Dirac} \\
\hline
 0~eV  & $1.03 \times 10^{-71}$ & $-2.83 \times 10^{-70}$ & $2.37 \times 10^{-69}$ & $-3.51 \times 10^{-70}$ & $1.26 \times 10^{-67}$ \\  
 
 0.1~eV & $2.34\times10^{-70}$ & $-3.75\times 10^{-69}$ & $1.63 \times 10^{-68}$  & $4.85 \times 10^{-69}$ & $1.26 \times 10^{-67}$ \\  
 \hline
\multicolumn{6}{c}{Majorana} \\
 \hline
 0~eV  & $3.18 \times 10^{-71}$ & $-6.24 \times 10^{-70}$ & $3.76 \times 10^{-69}$ & $1.94 \times 10^{-70}$ & $1.26 \times 10^{-67}$ \\  
 
 0.1~eV & $2.36\times10^{-70}$ & $-3.35\times 10^{-69}$ & $1.08 \times 10^{-68}$  & $2.32 \times 10^{-68}$ & $1.26 \times 10^{-67}$\\ 
 \hline

\end{tabular}
\label{Tab:ExpansionParameters}
\end{center}
\end{table*}
\end{document}